\documentclass[reprint,
superscriptaddress,
amsmath,amssymb,
aps,
prl,
longbibliography,
showpacs
]{revtex4-1}
\usepackage{graphicx}

\usepackage{color}

\begin{document}

\title{Engineering the Optical Spring via Intra-Cavity Optical-Parametric Amplification}
\author{Mikhail~Korobko}
\affiliation{Institut f\"ur Laserphysik und Zentrum f\"ur Optische Quantentechnologien, Universit\"at Hamburg, Luruper Chaussee 149, 22761 Hamburg, Germany}

\author{Farid Ya.~Khalili}
\email{khalili@phys.msu.ru}
\affiliation{Moscow State University, Department of Physics, Moscow 119992, Russia} 
\affiliation{Russian Quantum Center, Skolkovo 143025, Russia} 

\author{Roman~Schnabel}
\affiliation{Institut f\"ur Laserphysik und Zentrum f\"ur Optische Quantentechnologien, Universit\"at Hamburg, Luruper Chaussee 149, 22761 Hamburg, Germany}

\date{\today}

\begin{abstract}
The `optical spring' results from dynamical back-action and can be used to improve the sensitivity of cavity-enhanced gravitational-wave detectors. The effect occurs if an oscillation of the cavity length results in an oscillation of the intra-cavity light power having such a phase relation that the light's radiation pressure force amplifies the oscillation of the cavity length. 
Here, we analyse a Michelson interferometer whose optical-spring cavity includes an additional optical-parametric amplifier with adjustable phase. We find that the phase of the parametric pump field is a versatile parameter for shaping the interferometer's spectral density. 
\end{abstract}

%\pacs{03.67.Bg, 03.67.Hk, 42.50.Ex}
%\keywords{Quantum Physics}
\maketitle

\emph{Introduction} -- 
Electromagnetic dynamical back-action was first observed in radio-frequency systems and its existence predicted for optical Fabry-Perot cavities by Braginsky and his colleagues more that 50 years ago \cite{Braginsky1964, Braginsky1967}. The first proposal of using dynamical back-action to improve the sensitivity of laser-interferometric gravitational-wave detector was made in 1997, again by Braginsky and co-workers \cite{Braginsky1997}. The new scheme was called `optical bar', since the light's radiation pressure force rigidly connects two far separated mirrors, which are suspended as pendula but quasi-free otherwise. This way, a gravitational-wave signal is transformed into an acceleration of mirrors with respect to the local frame. The interferometric topologies that are considered in \cite{Braginsky1997} as well as in related work \cite{Khalili2002} are different from the Michelson topology having a balanced beam splitter, and were not experimentally realised so far. Recently, a more practical design was proposed \cite{Danilishin2006}.\\ 
The second proposal was made in 2002 by Buonanno and Chen \cite{Buonanno2002} and was called `optical spring'. It targets the sensitivity improvement of Michelson-type gravitational-wave detectors having a signal-recycling (SR) cavity \cite{Meers1988,Heinzel1998} or signal-extraction (SE) cavity, also called resonant-sideband extraction (RSE) \cite{Mizuno1993,Heinzel1996}. 
For the purpose of utilizing the optical spring in a Michelson interferometer operated on dark output port, these cavities need to be detuned from carrier light resonance. 
If the frequency of the carrier light is blue-detuned with respect to the cavity, the lower sidebands of phase modulations that are produced by gravitational waves and that are matching the detuning frequency get optically enhanced while the corresponding upper sidebands are suppressed. Due to energy conservation, the mechanical (pendulum) motion of the suspended mirror is enhanced \cite{Schliesser2006,Aspelmeyer2014}. The overall process corresponds to optomechanical parametric amplification and results in optical heating of the mechanical motion, i.e.~the opposite of optical cooling \cite{Metzger2004}. 
The radiation pressure of the light not only results in an optomechanical parametric amplification of the pendulum motion but also to an 
additional (optical) spring constant that increases the pendulum resonance frequency from typically 1\,Hz to an opto-mechanical resonance of up to about 100\,Hz. 
Around this frequency the mechanical response of the GW detector is significantly enhanced and its sensitivity improved. The frequency of the opto-mechanical resonance depends on the detuning and the optical power inside the arms of the detector. To further exploit the optical spring, it was  proposed to dynamically change the detuning by moving the SR/SE mirror in order to follow expected chirps of GW signals \cite{Meers1993,Simakov2014}. 
The optical spring was observed in several experiments \cite{Sheard2004,Miyakawa2006,Corbitt2006a,Corbitt2007,Corbitt2007a,Hossein-Zadeh2007,Aspelmeyer2014,Kelley2015,Sawadsky2015,Edgar2016,Singh2016,Gordon2017}. 
The gravitational-wave detectors GEO\,600 \cite{Affeldt2014}, Advanced LIGO \cite{Aasi2015}, Advanced Virgo \cite{Acernese2015}, and KAGRA \cite{Somiya2012} do use either SR or SE cavities, but so far have not yet employed the optical spring for a sensitivity enhancement due to the requirement of additional control techniques.\\
The conventional scheme for producing the optical spring does not use any additional parametric amplification of purely optical kind.  
Recent work, however, proposed complementing the SR/SE cavity with optical-parametric amplification \cite{Somiya2016} to allow for shifting up further the opto-mechanical resonance frequency without increasing the light power in the arms.

\begin{figure}[t!!!!!!]
	\centering
	\vspace{0mm}
	\includegraphics[width=0.35\textwidth]{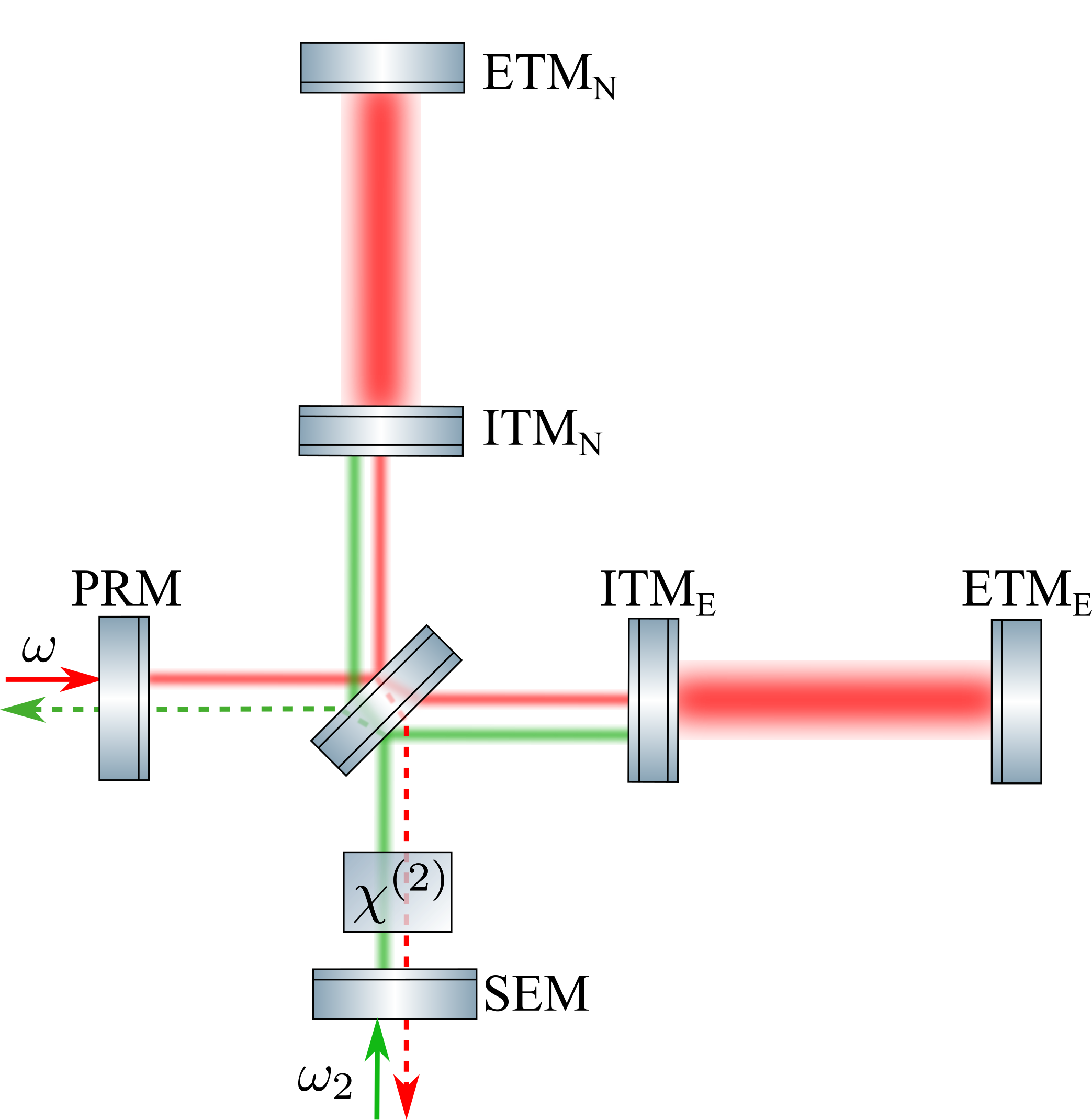}
		\vspace{-2mm}
	\caption{Schematic diagram of a proposed gravitational-wave detector. On top of the Advanced LIGO topology, consisting of arm resonators, a power-recycling mirror (PRM) and a signal-extraction mirror (SEM), a second-order nonlinear ($\chi^{(2)}$) crystal is placed in the SE cavity. The main carrier light at optical frequency $\omega$ is blue-detuned with respect to this cavity, but resonating in the arm cavities as well as in PR cavity. The second-order nonlinear crystal is pumped with a light field at frequency  $\omega_2 = 2\omega$ resulting in optical-parametric amplification (OPA) of light at $\omega$, including its quantum uncertainty. The pump field (displaced for better visibility) is also used to measure the differential motion of the ITMs. The two different wavelengths can be separated easily with dichroic beam splitters (not shown).
ITM$_{\rm N,E}$: input test mass in north and east arm, respectively. ETM$_{\rm N,E}$: end test mass.
}
	\label{fig:setup}
\end{figure}

In this work we extend the consideration in \cite{Somiya2016} and analyse the more general situation, in which not only the parametric gain is varied but also the angle of the amplified quadrature amplitude. The parametric gain relates to the \emph{intensity} of the second-harmonic pump field, whereas the angle relates to its \emph{phase}. In particular the last parameter can be quickly changed providing a new degree of freedom for realizing dynamical detuning of the optical spring properties. We consider the internal quantum noise squeezing that is accompanied with the optical-parametric amplification together with the one from the optomechanical parametric amplification and derive spectral densities.
Furthermore, we propose utilising the second-harmonic pump field to implement a `local readout' of the motion of the arm cavity input test masses (ITMs) \cite{Rehbein2007}, see Fig.\,\ref{fig:setup}. The local readout mitigates the unwanted effect of the optical spring, which is the rigid connection of the ITMs with their respective end test mass (ETM) at frequencies below the optical spring and a corresponding sensitivity loss at these frequencies.
 
\emph{The optomechanical system} --
The effects on the light field of both processes, the optical and the optomechanical parametric amplification, can be described in similar ways. 
Both result in a consecutive rotation of quadratures (determined by the phase of the pump for optical amplification and by the detuning of the SE cavity for the optomechanical one), squeezing (optical and ponderomotive correspondingly) and rotation again
{
 \cite{Harms2003, Danilishin2012}.
 }
Both types of parametric amplification should thus influence the optical spring also in similar ways. 
In this section we derive explicitly the optical spring in the case of additional optical-parametric amplification inside the SR/SE cavity and show the effect of the squeeze angle.

\begin{figure}[t!!!!!!!!!!!]
	\centering
	\vspace{0mm}
	\includegraphics[width=0.48\textwidth]{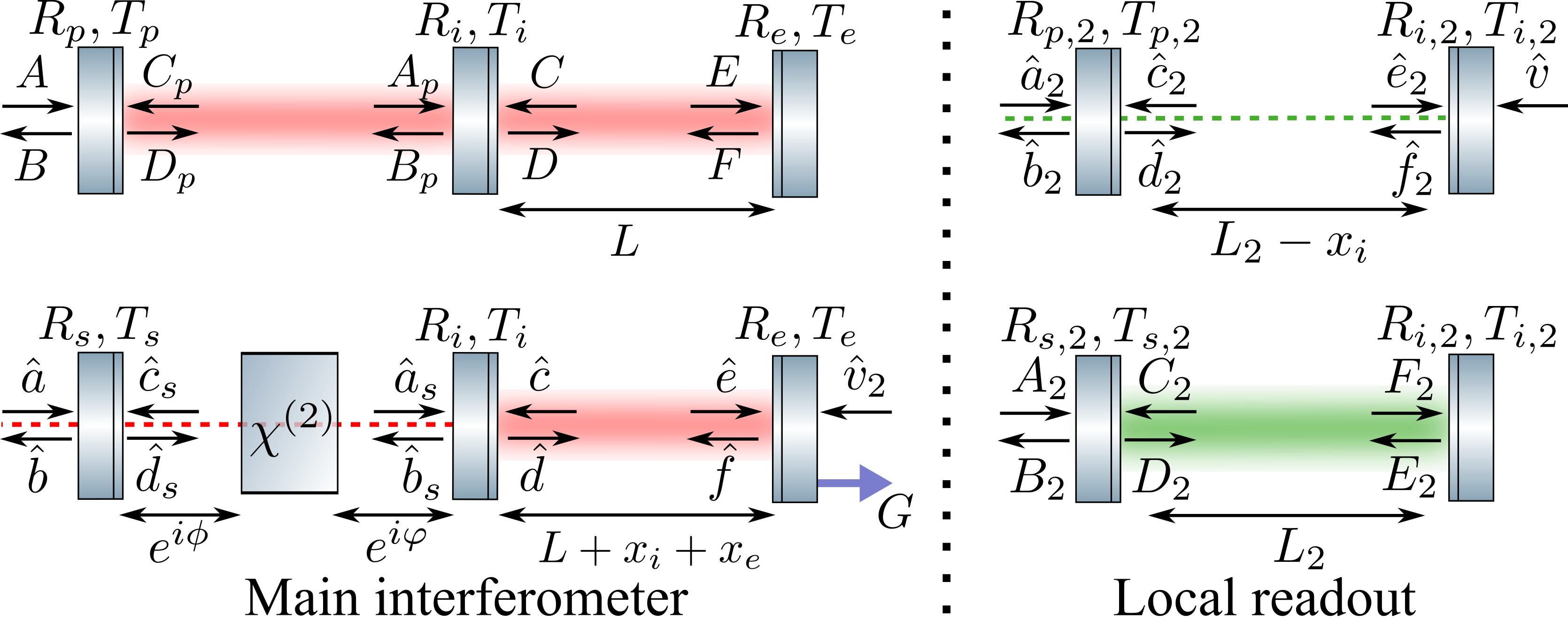}
		\vspace{-2mm}
	\caption{Notations of the optical fields for the PR cavity together with the common mode of the arm cavities at $\omega$ (top left), for the SE cavity together with the differential mode of the arm cavities at $\omega$ (bottom left), and respective parts of the interferometer in Fig.\,\ref{fig:setup} at $\omega_2$, which belongs to the local readout (top and bottom right). Operators are annihilation operators and denote complex amplitudes including their uncertainties. Capital letters $A$ to $F$ denote complex amplitudes whose uncertainties are irrelevant. Subscript `$s$': signal extraction; `$p$': power recycling; `$i$': input to arm cavity; `$e$': end of arm cavity; `$2$': optical frequency $\omega_2$. $R, T$: amplitude reflectivity and transmissivity of mirrors. $L$ is the average length of the arm resonators, ${{L}}_2$ is the relevant average length of the local read out, and $x_{i,e}$ represent their dynamical parts due to differential test mass motion. $\phi$ and $\varphi$ are additional phases accumulated by the light field inside the SE cavity due to the cavity detuning. The gravitational-wave signal (`G') corresponds to a differential change of the arm length $L$.
	}
	\label{fig:fields}
\end{figure}

We consider the dark port operation, at which all input light is retro-reflected from the interferometer. In this situation, the interferometer output field at the dark port contains the full information about the differential motion of the mirrors, and the output field at the bright port carries the full information about the common motion \cite{Caves1981}.\\ 
Let us first focus on the main interferometer that includes the arm cavities and is operated at optical frequency $\omega$. This interferometer includes the far mirrors (ETMs) whose differential distance from the rest of the interferometer is directly excited by the gravitational wave. (Depending on the polarisation and direction of propagation of the gravitational wave, also their common distance is changed, however, its measurement is not targeted by a Michelson interferometer.) 
With this in mind, we use an effective picture, where the interferometer is split into two separate cavity systems, coupled only via the displacement of the test mass mirrors \cite{Buonanno2003}. 
The first cavity system (Fig.\,\ref{fig:fields}, top left) corresponds to the common mode, whose modulation as well as its uncertainty are irrelevant for the signal-to-noise-ratio of a gravitational-wave signal in the differential mode. It is thus fully described by the classical carrier fields at frequency $\omega$. The second one corresponds to the differential mode at $\omega$ and requires a quantised description (Fig.\,\ref{fig:fields}, bottom left). The other two systems in Fig.\,\ref{fig:fields} (right) describe the short Michelson interferometer in Fig.\,\ref{fig:setup} that uses light at $\omega_2$ for pumping the optical-parametric process and for measuring the differential motion of the ITMs. This part of the interferometer is considered in the section \emph{local readout} further down.

{
The first mirrors of the main interferometer cavities (Fig.\,\ref{fig:fields}, left) }
are the power recycling (PRM) and signal extraction (SEM) correspondingly, and the middle (input, \textit{i}) and the end (\textit{e}) mirrors are combinations of ITM and ETM.
Then the differential motion of four mirrors can be defined as the motion of input and end mirrors in the effective cavity picture: 
\begin{multline}
\hat{x}_{-}(\Omega) = \left(x_{\rm ITM}^{(E)}(\Omega) + x_{\rm ETM}^{(E)}(\Omega)\right) -\\- \left(x_{\rm ITM}^{(N)} (\Omega)+ x_{\rm ETM}^{(N)}(\Omega)\right) = x_{i}(\Omega) + x_{e}(\Omega)\, .
\end{multline}

Relative to the beam splitter only the far mirrors are accelerated due to the gravitational wave force $G$, as the input mirrors are so close to the beam splitter that the effect can be neglected.
On top, all mirrors are accelerated by the light's radiation pressure force $F^{ba}_{(i,e,2)}$, which is proportional to the power of the light shining on the mirror and which we call back-action.
\begin{align}\label{eq:mech_motion}
&\hat{x}_i(\Omega) = \chi_i(\Omega) \left[ F^{ba}_{i} - F^{ba}_2 \right]\, ,\\
&\hat{x}_e(\Omega) = \chi_e(\Omega) \left[ F^{ba}_{e} + G\right]\, ,
\end{align}
where 
{
$\chi_{i, e} = [-m\Omega^2]^{-1}$ 
}
are mechanical susceptibilities of the input and end mirrors, that we assume here to be identical quasi-free masses of mass $m$. 
The input mirror is driven by two different optical forces, due to the additional back-action $F^{ba}_2$ from the second harmonic pump field.

\emph{The optical spring} -- 
When the cavity is detuned from resonance, the back-action force has a position-dependent dynamical part, causing the optical spring effect. 
We thus split the force into two contributions -- the fluctuating part due to the quantum uncertainty of the light's amplitude quadrature, and the optical spring force $F^{ba} = F^{fl}(\Omega) - \mathcal{K}(\Omega) x(\Omega)$, where $\mathcal{K}(\Omega)$ is the optical spring constant, also called \text{optical rigidity} .

We calculate the optical rigidity in the single-mode approximation, where the back-action on the input and end mirrors are identical, yielding
\begin{equation}\label{eq:ba}
F^{ba}_{i, e}(\Omega)  =  F_{fl}(\Omega) - \mathcal{K}(\Omega) x_-(\Omega)\, .
\end{equation}
The single-mode approximation \cite{Danilishin2012} further involves 
\textit{(i)} the sideband frequency and the arm cavity detuning being much smaller than the cavity free spectral range $\Omega, \delta_{a}\ll c/L$, with $L$ being the arm cavity length, and $c$ the speed of light, and 
\textit{(ii)} the transmissivity $T_{i,e}$ of mirrors being small, so that we can make a Taylor expansion $R_{i,e} \approx 1-T_{i,e}^2/2$.
The single-mode approximation enables us to introduce an effective linewidth $\gamma$ and the detuning of the SE cavity $\delta_s$ as well as the normalised optical parametric gain (per cavity round trip) $\Gamma$ in the following way
\begin{align}
\gamma &= \frac{\gamma_a T_s}{D_0}\, ,\\
\delta_s &= \frac{2 \gamma_a R_s}{D_0} (\cosh 2q \cos 2\phi \sin 2\varphi + \sin 2\phi \cos 2\varphi)\, ,\\
\Gamma &= \frac{2\gamma_a R_s \sinh 2q \cos 2\phi}{D_0}\, , \; {\rm with} 
\end{align}
\begin{multline}
D_0 = 1 + 2 R_s (\cosh 2 q \cos 2\phi \cos 2\varphi -\\ - \sin 2\phi \sin 2\varphi) + R_s^2\, ,
\end{multline}
where $\gamma_a = c (T_i^2 + T_e^2)/(4 L)$ is the linewidth of the arm cavity, $\phi$ and $\varphi$ are additional phases accumulated by the light field inside the SE cavity due to the cavity detuning and $q$ is a squeeze factor on the single pass through the optical-parametric amplifier.

We find that the optical parametric gain $\Gamma$ influences the total detuning of the interferometer $\delta_{\rm eff}$ as well as the light power associated with the optical spring $J_{\rm eff}$
\begin{align}
\delta_{\rm eff} &= \sqrt{\delta^2 - \Gamma^2}\, ,\\
J_{\rm eff}& = J (\delta - \Gamma \sin 2 \theta) / \delta_{\rm eff}\, ,
\end{align}
where $\delta=\delta_a + \delta_s$ with $\delta_a$ the arm cavity detuning. $\theta$ is the phase of the optical-parametric amplification (the squeeze angle), and $J = 4 \omega I_c/(m c L)$ the normalised optical power with $I_c$ the power circulating in the arm cavities.

Given these definitions the optical rigidity $\mathcal{K}(\Omega)$ is found to be
\begin{equation}
\mathcal{K}(\Omega) = \frac{m J (\delta - \Gamma \sin 2\theta)}{(\gamma - i\Omega)^2 + \delta^2 - \Gamma^2} = \frac{m J_{\rm eff} \delta_{\rm eff}}{(\gamma - i\Omega)^2 + \delta_{\rm eff}^2}.
\end{equation}

The optical spring, enhanced by the optical-parametric amplifier, has several important properties.

\textit{First}, the maximal enhancement of the optical spring due to the internal squeezing is achieved if $\theta = -\pi/4$ (for $\delta > 0$) yielding $J<J_{\rm eff}\propto J e^{2q}$. In this case, for instance, 3\,dB of intra-cavity squeezing modifies the optical spring in the same way as doubling the optical power.

\begin{figure}[bt]
	\centering
	\vspace{0mm}
	\includegraphics[width=0.46\textwidth]{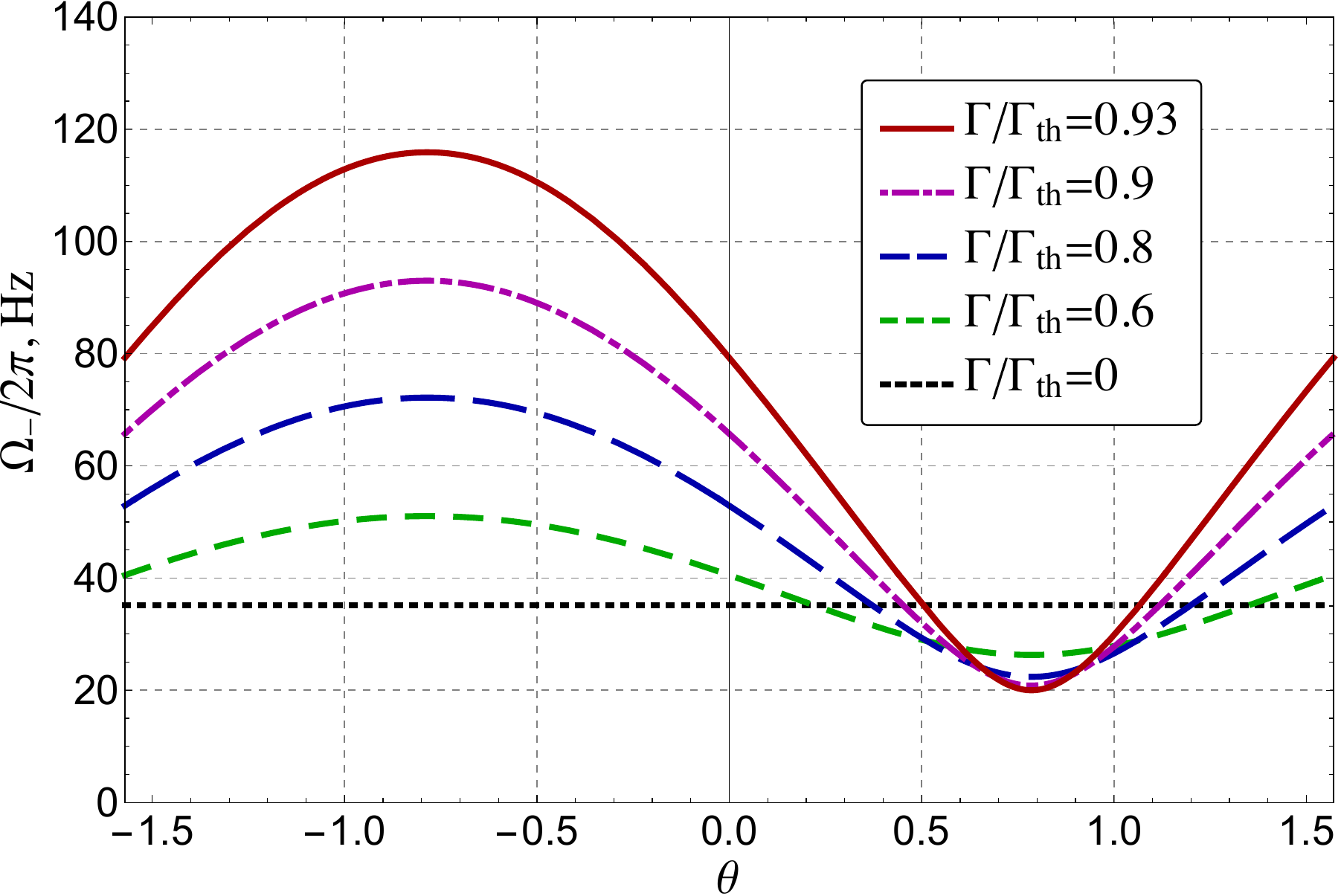}
		\vspace{-2mm}
	\caption{Optomechanical frequency $\Omega_-/2\pi$ as a function of squeeze angle $\theta$ for different values of parametrical gain $\Gamma$ relative to the threshold value $\Gamma_{\rm th}$. The detuning is fixed $\delta/2 \pi = 580$\,Hz, and all other parameters correspond to the AdvLIGO parameters \cite{Aasi2015}. }
	\label{fig:3}
\end{figure}

\textit{Second}, we note that the internal squeezing changes the dynamics and stability of the system.
The characteristic equation for the optomechanical motion is
\begin{multline}\label{eq:ch_eq}
\Omega^4 + 2 i\Omega^3\gamma + \Omega^2 (\Gamma^2 - \gamma^2 - \delta^2) + \\ + J(\delta - \Gamma \sin 2\theta) = 0.
\end{multline}
The resonances can be found in the perturbative way by expanding the roots of Eq.~\eqref{eq:ch_eq} in powers of $\gamma$.
Then zeroth order of expansion gives two positive roots:
\begin{align}
&\Omega_{\pm}^{(0)} = \sqrt{\frac{\delta_{\rm eff}^2}{2} \pm \sqrt{\frac{\delta_{\rm eff}^4}{4}-J_{\rm eff}\delta_{\rm eff}}},
\end{align}
where $\Omega_{-}^{(0)}$ corresponds to the shifted mechanical resonance, and $\Omega_{+}^{(0)}$ to the optical resonance.
In the absence of optomechanical coupling, $J_{\rm eff} = 0$, mechanical resonance is at zero, 
which corresponds to our assumption of having quasi-free masses, 
and $\Omega_+^{(0)} = \delta_{\rm eff} = \sqrt{\delta^2 - \Gamma^2}$.
With increased effective power the mechanical resonance shifts to higher frequencies, and the optical one gets reduced, until, within the approximation used,  these resonances become equal at the critical power $J_{\rm eff}^{\rm (c)} = \delta_{\rm eff}^3/4$ 
\cite{Khalili2001}.  
Note that the absence of optomechanical coupling ($J_{\rm eff} = 0$) can be due to zero power ($J = 0$) or if the condition $\Gamma \sin 2\theta = \delta$ holds. Generally, the effective power can be changed by tuning the squeeze angle, without affecting neither light power nor squeeze factor, see Fig.\,\ref{fig:3}. 

\begin{figure}[bt]
	\centering
	\vspace{0mm}
	\includegraphics[width=0.46\textwidth]{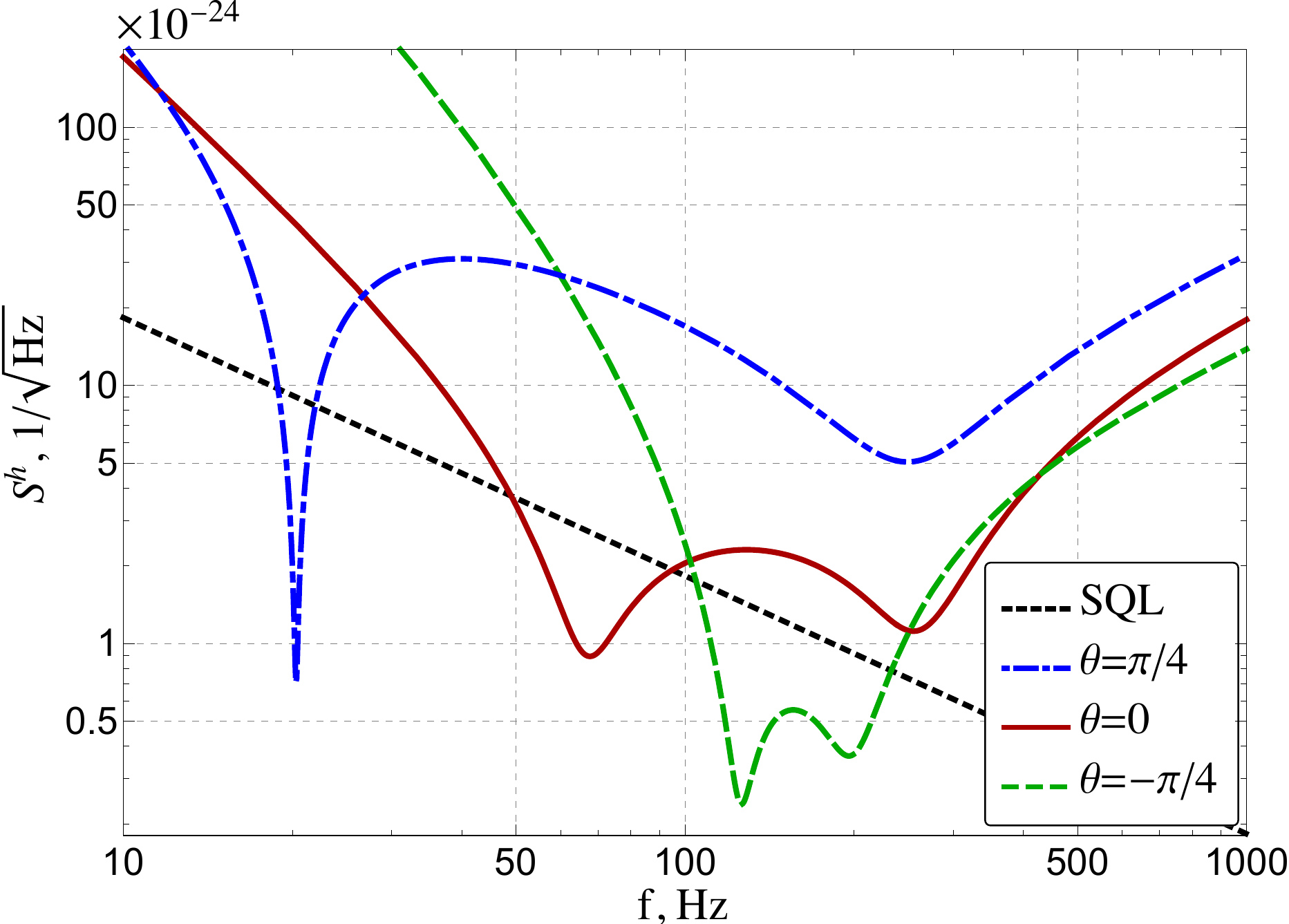}
		\vspace{-2mm}
	\caption{Example of tuning of the sensitivity by changing only the squeeze angle of the intra-cavity amplifier. Three plots correspond to three different values of the angle, while all the parameters are fixed to be equal to the current AdvLIGO design parameters, with $\delta/2 \pi = 580$\,Hz  and $\Gamma/\Gamma_{\rm th} = 0.93$. On the plot one can see how depending on the squeeze angle different frequencies obtain the maximal sensitivity, covering a vast frequency band, that opens a possibility for dynamical tuning. This demonstrates the possibility to engineer the optical spring by changing only the squeeze angle. Note that the generally poor sensitivity at low frequencies can be improved by an additional local read out of the differential motion of the near test masses \cite{Rehbein2007}.}
	\label{fig:4}
\end{figure}

We propose thus to use this feature for \textit{dynamical tuning} of the interferometer response to the GW signal.
This way the mechanical resonance is changing adaptively to match the chirp GW signal, see Fig.\,\ref{fig:4}.
Tuning of the squeeze angle can be done in a straightforward way by tuning the phase of the second harmonic pump, 
e.g.~by transmission through a fast electro-optical modulator.
We note that the tuning speed is ultimately limited by the decay rate of the pump light's cavity.

\textit{Third}, using Routh-Hurwitz' criterion \cite{gradshteyn2007} one can show, that the system is always unstable, in the same way the optical spring without internal squeezing is \cite{Buonanno2001, Buonanno2002, Buonanno2003}. 
The mechanical system can be stabilised via active feedback \cite{Buonanno2002}. The optical-parametric process leads to an additional stability condition that has to be satisfied: $\Gamma^2 < \Gamma_{\rm th}^2 = \gamma^2 + \delta^2$, which is a threshold for optical-parametric instability.

\emph{Local readout} --
The optical spring can be efficiently combined with the \textit{local readout} \cite{Rehbein2007}.
The idea of the local readout is based on that of the \textit{optical bar}, proposed by Braginsky and co-authors.
At frequencies below the optomechanical resonance, the ITMs and ETMs are connected in a rigid way via dynamical backaction. 
For this reason, at these low frequencies, the motion of the ETMs due to a gravitational wave  causes an identical motion of the ITMs, which reduces the sensitivity at these frequencies.
The idea behind the local readout is to measure the motion of the ITMs locally and to use this information in the data processing. This way the sensitivity can be greatly improved at low frequencies.
Here we propose to use the second harmonic pump of the OPA to sense the local motion of the ITM.
In this case the dark port of a second small Michelson interferometer, formed by ITMs, coincide with the bright port of the main interferometer, see Fig.\,\ref{fig:setup} and \ref{fig:fields}. 
Both outputs have to be measured with balanced homodyne detectors with an optimal homodyne angle, and then combined in an optimal way.
For the local readout it is important to take the motion of the central beamsplitter into account, as the arm length of the small interferometer is rather short and arm cavities are not used. \\
Using the second harmonic pump instead of a nearby wavelength \cite{Rehbein2007} most likely eases the fabrication of the optical components that are part of both interferometers. The large wavelength separation allows to manufacture a highly reflective optical coating on the ITMs for the local readout beam, while keeping a moderate transmissivity for the main beam to allow for overcoupling of the input light. 
The detection of the two beams can be done independently in efficient way, by separating the beams with dichroic beam splitters, which avoids using polarisation optics or filter cavities, as proposed in \cite{Rehbein2007}.

\emph{Summary and outlook} --
In this work, we analyse the optical spring that is created by the `conventional' optomechanical parametric amplification inside a detuned cavity in combination with intra-cavity optical parametric amplification (OPA). Modifying the spectral density of the \emph{conventional} optical-spring GW detector requires modifying the light power in the interferometer arms and/or changing the length of the signal recycling/signal extraction cavity. We find that the same modification is achieved for the \emph{OPA-enhanced} optical spring by changing the power of the OPA pump light and/or by changing the phase of the pump light.  
We conjecture that the latter parameters can be changed during an observational run of the GW detector more easily. 
We also conjecture that in practice tuning and controlling these parameters should be possible with higher bandwidth, which is, however, ultimately limited by the pump light's cavity decay rate. 
Our analysis includes the effect of optical-parametric quantum noise squeezing inside the SR/SE cavity, which is also called \emph{internal} squeezing \cite{Korobko2017}. It is qualitatively different from external squeezing \cite{Caves1981,Schnabel2010}, which has been exploited in the GW detector GEO\,600 since 2010 \cite{LSC2011}. Our analysis is in principle extendable to external squeezing, which was investigated in detail for the conventional optical spring in \cite{Harms2003}.\\
We propose exploiting the second-harmonic pump field of the OPA for a local readout to increase the sensitivity at low frequencies. 
The exploitation of the pump field as the second carrier field might also be interesting for other schemes such as the double optical spring \cite{Rehbein2008} or multi-carrier configurations \cite{Khalili2011,Voronchev2012a,Korobko2015}.\\ 
Our work extends both aspects of the optical bar, the optical spring as well as the local readout, towards gravitational-wave detectors with intra-cavity parametric amplification, and shows that such an approach allows versatile engineering of GW detector sensitivities.\\

\emph{Acknowledgments} --
The work of MK and RS was supported by the Deutsche Forschungsgemeinschaft (DFG) (SCHN~757/6-1).
{
The work of FK was supported by LIGO NSF Grant PHY-1305863 and by Russian Foundation for Basic Research Grants 16-52-10069 and 16-52-12031.
}

\appendix
\section{Appendix A: Input-output relations}
It is helpful to consider the input-output relations of our opto-mechanical system in the `two-photon formalism' \cite{Caves1985a,Schumaker1985a}, where the amplitude and phase quadrature amplitudes $\hat{a}^{c}$ and $\hat{a}^{s}$ of the modulation field at frequency $\Omega$ are linked to the optical fields $\hat{a}(\omega \pm \Omega)$ via 
\begin{align}
\hat{a}^{c}(\Omega) = \frac{\hat{a}(\omega +  \Omega) + \hat{a}^\dag(\omega - \Omega))}{\sqrt{2}}\, , \\ 
\hat{a}^{s}(\Omega) = \frac{ \hat{a}(\omega + \Omega) - \hat{a}^\dag(\omega - \Omega)}{i\sqrt{2}}\, .
\end{align}
Based on these quadrature amplitudes we define the vector $\mathbf{\hat{a}}(\Omega) = \{\hat{a}^c(\omega), \hat{a}^s(\Omega)\}^{\rm T}$.

The signal recycling cavity rotates the quadratures due to it's detuning and squeezes and rotates additionally due to intra-cavity optical-parametric amplification. The phase shift due to the cavity length can be neglected since the cavity length is much shorter than the wavelengths of the sideband modulations considered here. The effect of the signal recycling cavity can be described as a set of rotations and squeezing operations:
\begin{align}
\mathbf{\hat{a}}_s &= \mathbb{O}(\varphi)\mathbb{O}(\theta)\mathbb{S}\mathbb{O}^\dag(\theta)\mathbb{O}(\phi) \mathbf{\hat{d}}_s\, ,\\
\mathbf{\hat{b}}_s &= -R_i \mathbf{\hat{a}}_s + T_i \mathbf{\hat{c}},\\
\mathbf{\hat{c}}_s & =\mathbb{O}(\phi)\mathbb{O}(\theta)\mathbb{S}\mathbb{O}^\dag(\theta)\mathbb{O}(\varphi) \mathbf{\hat{b}}_s\, ,\\
\mathbf{\hat{d}}_s &=T_s \mathbf{\hat{a}} + R_s \mathbf{\hat{c}}_s,
\end{align}
where we denote the amplitude reflectivity and transmissivity of the signal recycling and input mirrors by $R_{s, i}, T_{s,i }$ and the phase delay due to the cavity detuning before and after the crystal by $\phi, \varphi$, see Fig.\,\ref{fig:fields}.
We now introduce the squeeze angle $\theta$ and the rotation matrix 
\begin{equation}
\forall \phi, \quad \mathbb{O}(\phi) = 
\begin{bmatrix}
\cos \phi & -\sin \phi \\ \cos \phi & \sin \phi
\end{bmatrix},
\end{equation}
\begin{equation}
\mathbb{Y} = \mathbb{O}(\pi/2) = 
\begin{bmatrix}
0 & -1 \\ 1 & 0
\end{bmatrix}
\end{equation}
and squeezing matrix 
\begin{equation}
\mathbb{S} = 
\begin{bmatrix}
e^{q} & 0 \\ 0 & e^{-q}
\end{bmatrix},
\end{equation}
with $q$ being the single-pass squeeze factor.

For the arm cavity the corresponding set of equations reads
\begin{align}
\mathbf{\hat{b}} & = -R_s \mathbf{\hat{a}} + T_s \mathbf{\hat{c}}_s\, ,\\
\mathbf{\hat{d}} & = R_i \mathbf{\hat{c}} + T_i \mathbf{\hat{a}}\, ,\\
\mathbf{\hat{c}} &= \mathbb{O}(\delta_a \tau_a)\mathbf{\hat{f}} e^{i\Omega \tau_a}\, , \\
\mathbf{\hat{e}} &= \mathbb{O}(\delta_a \tau_a)\mathbf{\hat{d}}e^{i\Omega \tau_a}\, ,\\
\mathbf{\hat{f}} & = R_e\mathbf{\hat{e}} + T_e \mathbf{\hat{v}}   + 2 k \mathbb{O}(\pi/2)\mathbf{E} \hat{x}_-(\Omega) \, , %
\end{align}
where $k = \omega/c$ is the wave vector of the main field, $\delta_a$ is the arm cavity detuning  and $\tau_a = L/c$  is the propagation time with $L$ being the length of the arm cavity, and $c$ the speed of light.
The field $\mathbf{E}$ corresponds to the classical amplitude of the field impinging on the end mirror. This set of equations can be resolved for the outgoing field $\mathbf{\hat{b}}$ and intra-cavity fields $\mathbf{\hat{c}},\mathbf{\hat{d}},\mathbf{\hat{e}},\mathbf{\hat{f}}$.

We find the solution to these equations, first for the complex transmissivity and reflectivity of the signal recycling cavity
\begin{align}
& \mathbf{\hat{b}}_s = \mathbb{D}_b \left[ -R_i T_s\mathbb{M}[\varphi, \phi] \mathbf{\hat{a}} + T_i \mathbf{\hat{c}}\right]\, ,\\
& \mathbf{\hat{d}}_s = \mathbb{D}_d \left[ R_s T_i\mathbb{M}[\phi, \varphi] \mathbf{\hat{c}} + T_s \mathbf{\hat{a}}\right]\, ,\\
& \mathbf{\hat{a}}_s = \mathbb{M}[\varphi,\phi]\mathbb{D}_d \left[ R_s T_i\mathbb{M}[\phi, \varphi] \mathbf{\hat{c}} + T_s \mathbf{\hat{a}}\right]\, ,\\
& \mathbf{\hat{c}}_s = \mathbb{M}[\phi,\varphi]\mathbb{D}_b \left[ -R_i T_s\mathbb{M}[\varphi, \phi] \mathbf{\hat{a}} + T_i \mathbf{\hat{c}}\right]\, ,
\end{align}
where we defined
\begin{align}
&\mathbb{M}[\phi,\psi] = \mathbb{O}(\phi)\mathbb{O}(\theta)\mathbb{S}\mathbb{O}^\dag(\theta)\mathbb{O}(\psi), \forall \phi, \psi\, ,\\
&\mathbb{D}_b = \left( \mathbb{I} + R_i R_s  \mathbb{M}[\varphi,\phi]\mathbb{M}[\phi,\varphi]\right)^{-1}\, ,\\
&\mathbb{D}_d = \left( \mathbb{I} + R_i R_s  \mathbb{M}[\phi,\varphi]\mathbb{M}[\varphi,\phi]\right)^{-1}\, .
\end{align}
That provides the solution for the signal extraction cavity
\begin{align}
&\mathbf{\hat{b}} = -\mathbb{R}_b \mathbf{\hat{a}} + \mathbb{T}_b \mathbf{\hat{c}}\, , \\
&\mathbf{\hat{d}} = -\mathbb{R}_d \mathbf{\hat{c}} + \mathbb{T}_d \mathbf{\hat{a}}\, ,
\end{align}
where 
\begin{align}
&\mathbb{R}_b = R_s + R_i T_s^2 \mathbb{M}[\phi,\varphi]\mathbb{D}_b \mathbb{M}[\varphi,\phi]\, ,\\
&\mathbb{R}_d = R_i + R_s T_i^2 \mathbb{M}[\varphi,\phi]\mathbb{D}_d \mathbb{M}[\phi,\varphi]\, ,\\
&\mathbb{T}_b = T_i T_s \mathbb{M}[\phi,\varphi]\mathbb{D}_b\, , \\
&\mathbb{T}_d = T_i T_s \mathbb{M}[\varphi,\phi]\mathbb{D}_d\, . 
\end{align}
Now we can derive the fields for the arm cavity yielding
\begin{align}
&\begin{aligned}
\mathbf{\hat{c}} = & R_e \mathbb{D}_c \mathbb{O}(\delta_a \tau_a)^2 \mathbb{T}_d \mathbf{\hat{a}} e^{2 i \Omega \tau_a} + \\  &+  T_e \mathbb{D}_c\mathbb{O}(\delta_a \tau_a)\mathbf{\hat{v}} e^{ i \Omega \tau_a}\\ &+ 2 k \mathbb{D}_c \mathbb{O}(\delta_a \tau_a) \mathbb{Y}\mathbf{E} \hat{x}_-(\Omega) e^{i\Omega\tau_a}
\end{aligned}\\
&\begin{aligned}
\mathbf{\hat{e}} = & \mathbb{D}_e \mathbb{O}(\delta_a \tau_a) \mathbb{T}_d \mathbf{\hat{a}} e^{i \Omega \tau_a} +\\ &+ T_e \mathbb{D}_e \mathbb{O}(\delta_a \tau_a) \mathbb{R}_d \mathbb{O}(\delta_a \tau_a) \mathbf{\hat{v}} e^{2 i \Omega \tau_a} +\\ & +2 k \mathbb{D}_e \mathbb{O}(\delta_a \tau_a) \mathbb{R}_d \mathbb{O}(\delta_a \tau_a) \mathbb{Y}\mathbf{E} \hat{x}_-(\Omega) e^{2i\Omega\tau_a}
\end{aligned}
\end{align}
where
\begin{align}
\mathbb{D}_c &=  \left( \mathbb{I} - R_e\mathbb{O}(\delta_a \tau_a)^2 \mathbb{R}_d e^{2i\Omega \tau_a}\right)^{-1}\\
\mathbb{D}_e &=  \left( \mathbb{I} - R_e\mathbb{O}(\delta_a \tau_a) \mathbb{R}_d \mathbb{O}(\delta_a \tau_a) e^{2i\Omega \tau_a}\right)^{-1}
\end{align}
Finally, we find the outgoing field to be
\begin{align}
\mathbf{\hat{b}} = & \left(-\mathbb{R}_b + R_e\mathbb{T}_b \mathbb{D}_c\mathbb{O}(\delta_a \tau_a)^2 \mathbb{T}_d e^{2i\Omega\tau_a}\right) a + \\  &+T_e \mathbb{T}_b\mathbb{D}_c\mathbb{O}(\delta_a \tau_a)\mathbf{\hat{v}} e^{ i \Omega \tau_a} +\\  &+ 2 k \mathbb{T}_b\mathbb{D}_c \mathbb{O}(\delta_a \tau_a) \mathbb{Y}\mathbf{E} \hat{x}_-(\Omega) e^{i\Omega\tau_a} = \\ &= -\mathbb{R} \mathbf{\hat{a}}  + \mathbb{T}\mathbf{\hat{v}} + \mathbb{Z} \hat{x}_-(\Omega)
\end{align}

\section{Appendix B: Radiation pressure}
The radiation pressure force acting on the mirrors has three contributions. First, there is a constant force due to the classical high-power optical field. It induces a constant shift of the mirror, which can be compensated with classical feedback.
Second, there is a dynamical classical part, which is amplified by opto-mechanical parametric amplification and which belongs to the optical spring, and third a fluctuating force due to the uncertainty in the amplitude quadrature of the light.
The latter corresponds to the quantum back-action force of the carrier light and can be written for the input and end mirrors as
\begin{align}
&F^{ba}_{i} = \hbar k (\mathbf{C}^\dag \mathbf{\hat{c}}(\Omega) + \mathbf{D}^\dag \mathbf{\hat{d}}(\Omega))\, ,\\
&F^{ba}_{e} = \hbar k (\mathbf{E}^\dag \mathbf{\hat{e}}(\Omega) + \mathbf{F}^\dag \mathbf{\hat{f}}(\Omega))\, .
\end{align}
In the single-mode approximation, these two forces become equal and read
\begin{equation}
F^{ba}_{i, e}(\Omega)  =  2\hbar k \mathbf{E}^\dag \mathbf{\hat{e}}(\Omega)= F_{fl}(\Omega) - \mathcal{K}(\Omega) x_-(\Omega)\, ,
\end{equation}
where
\begin{align}
&\begin{aligned}
F^{fl}(\Omega) = 2\hbar k \mathbf{E}^\dag \mathbb{D}_e & \mathbb{O}(\delta_a \tau_a) e^{i \Omega \tau_a} \left(\mathbb{T}_d \mathbf{\hat{a}} + \right. \\ & \left.+ T_e \mathbb{R}_d \mathbb{O}(\delta_a \tau_a) \mathbf{\hat{v}} e^{i \Omega \tau_a} \right)
\end{aligned}\\
&\mathcal{K}(\Omega) = -4\hbar k^2 \mathbf{E}^\dag \mathbb{D}_e \mathbb{O}(\delta_a \tau_a) \mathbb{R}_d \mathbb{O}(\delta_a \tau_a) \mathbb{Y}\mathbf{E} \hat{x}_-(\Omega) e^{2i\Omega\tau_a}
\end{align}
Ignoring the effect of the second-harmonic beam, we get for the differential motion
\begin{equation}
\hat{x}_-(\Omega) = \chi(\Omega) \left[ F^{fl} - \mathcal{K}(\Omega) x_-(\Omega) \right]\, ,
\end{equation}
which allows us to introduce an effective susceptibility:
\begin{equation}
\chi_{\rm eff}(\Omega) = (\chi^{-1} + \mathcal{K}(\Omega))^{-1}\, ,
\end{equation}
such that $x_-(\Omega) = \chi_{\rm eff}(\Omega) F^{fl}(\Omega)$.

\section{Appendix C: Detection}
The balanced homodyne detection on the output $\mathbf{\hat{b}}$ at homodyne angle $\zeta$ provides the values
\begin{multline}
y(\Omega) = \begin{bmatrix}
\cos \zeta & \sin \zeta
\end{bmatrix}
\mathbf{\hat{b}}(\Omega) = \mathbb{H}^{\rm T}\mathbf{\hat{b}}(\Omega) = \\= -\mathbb{H}^{\rm T}\mathbb{R}\mathbf{\hat{a}}  + \mathbb{H}^{\rm T}\mathbb{T}\mathbf{\hat{v}}+ \mathbb{H}^{\rm T}\mathbb{Z} \hat{x}_-(\Omega)
\end{multline}
which we renormalise to the differential mirror displacement
\begin{multline}
\tilde{y} = \frac{-\mathbb{H}^{\rm T}\mathbb{R}\mathbf{\hat{a}}  + \mathbb{H}^{\rm T}\mathbb{T}\mathbf{\hat{v}}}{\mathbb{H}^{\rm T}\mathbb{Z} }  + \hat{x}_-(\Omega) =\\ = -\frac{-\mathbb{H}^{\rm T}\mathbb{R}\mathbf{\hat{a}}  + \mathbb{H}^{\rm T}\mathbb{T}\mathbf{\hat{v}}}{\mathbb{H}^{\rm T}\mathbb{Z} }  + \chi_{\rm eff} F^{fl} 
\end{multline}
From this we get the spectral density for this output
\begin{multline}
S_x(\Omega) = S_{xx}(\Omega) + 2 {\rm Re}[\chi^*_{\rm eff}(\Omega) S_{xF}(\Omega)] + \\+ |\chi_{\rm eff}(\Omega)|^2 S_{FF}(\Omega), 
\end{multline}
where
\begin{align}
&S_{xx} = \frac{\mathbb{H}^{\rm T}(\mathbb{R}\mathbb{R}^\dag + \mathbb{T}\mathbb{T}^\dag)\mathbb{H}}{|\mathbb{H}^{\rm T}\mathbb{Z}|^2},\\
&\begin{aligned} 
S_{FF} = 4 \hbar^2 k^2 & \mathbf{E}^\dag \mathbb{D}_e \mathbb{O}(\delta_a \tau_a) \times \\  &\times \left(\mathbb{T}_d  \mathbb{T}_d^\dag  + T_e^2 \mathbb{R}_d  \mathbb{R}_d^\dag\right) \mathbb{O}^\dag(\delta_a \tau_a) \mathbb{D}_e^\dag \mathbf{E},
\end{aligned}\\
&\begin{aligned}
S_{xF} &= \frac{2\hbar k}{\mathbb{H}^{\rm T}\mathbb{Z}} \left(-\mathbb{H}^{\rm T}\mathbb{R} \mathbb{T}_d^\dag \mathbb{O}^\dag(\delta_a \tau_a) \mathbb{D}_e^\dag \mathbf{E} e^{-i\Omega \tau_a} +\right. \\ & + \left. T_e \mathbb{H}^{\rm T}\mathbb{T}  \mathbb{O}^\dag(\delta_a \tau_a) \mathbb{R}_d \mathbb{O}^\dag(\delta_a \tau_a) \mathbb{D}_e^\dag \mathbf{E} e^{-2i\Omega \tau_a} \vphantom{ \frac{2\hbar k}{\mathbb{H}^{\rm T}\mathbb{Z}}} \right).
\end{aligned}
\end{align}
Finally we normalise the spectral density to the gravitational-wave strain yielding
\begin{equation}
S_h(\Omega) = S_x(\Omega) \frac{4}{m^2 L^2 \Omega^4 |\chi_{\rm eff}(\Omega)|^2}\, .
\end{equation}

\bibliography{library-RS}

\end{document}